\documentclass[prl,aps,twocolumn,superscriptaddress,preprintnumbers,amsmath,amssymb,showpacs]{revtex4}
\usepackage{graphicx}
\usepackage{dcolumn}
\usepackage{bm}
\usepackage{longtable}
\begin{document}
\title{Coexistence of Bound and Virtual-bound States in Shallow-core to Valence Spectroscopies}
\author{Subhra Sen Gupta}
\email{subhra@phas.ubc.ca}
\affiliation{Department of Physics and
Astronomy, University of British Columbia, Vancouver, BC V6T 1Z1,
Canada.}
\author{J. A. Bradley}
\affiliation{Department of Physics, University of Washington,
Seattle, Washington 98105, USA.}
\author{M. W. Haverkort}
\affiliation{Max Planck Institute for Solid State Research,
Heisenbergstra{\ss}e 1, D-70569 Stuttgart , Germany.}
\author{G. T. Seidler}
\affiliation{Department of Physics, University of Washington,
Seattle, Washington 98105, USA.}
\author{A. Tanaka}
\affiliation{ Department of Quantum Matter, ADSM, Hiroshima
University, Higashi-Hiroshima 739-8530, Japan.}
\author{G. A. Sawatzky}
\affiliation{Department of Physics and
Astronomy, University of British Columbia, Vancouver, BC V6T 1Z1,
Canada.}

\date{\today}

\begin{abstract}

We develop the theory for shallow-core to valence excitations when
the multiplet spread is larger than the core-hole attraction,
\emph{e.g.}, if the core and valence orbitals have the same
principal quantum number. This results in a cross-over from bound to
virtual-bound excited states with increasing energy and in large
differences between dipole and high-order multipole transitions, as
observed in inelastic x-ray scattering. The theory is important to
obtain ground state information from x-ray spectroscopies of
strongly correlated transition metal, rare-earth and actinide
systems.

\end{abstract}










\maketitle

The actinides and their compounds are attracting serious attention
from the condensed matter community due to their exotic
properties~\cite{RMP-Moore}. Examples of these are the extremely
rich phase diagram of Plutonium~\cite{Hecker-Pu} and the unsolved
``hidden order" transition~\cite{URu2Si2} in URu$_{2}$Si$_{2}$.
Their properties interpolate between more itinerant $3d$ transition
metal (TM) systems and more localized $4d$ TM
compounds~\cite{vanderMarel}, and exhibit strong interplay between
spin, charge, orbital, and lattice degrees of freedom. Thus,
versatile experimental techniques are needed to unravel the physics
operative in these systems. Core-level spectroscopies, like x-ray
absorption spectroscopy (XAS), have been extremely successful in
providing information regarding the ground state of TM and
rare-earth (RE) systems, relying largely on theoretical
interpretations based on local correlated models with full multiplet
effects~\cite{degroot-kotani-book} and atomic selection rules.

The success of such local multiplet models relies strongly on two
facts : (1) their electronic structure is largely governed by local
correlation physics and point group symmetry, and (2) the final
state core-hole strongly binds the extra $d$ or $f$ electron,  so
that \emph{all} core-valence multiplets (CVM) form \emph{excitonic
bound states}. However, core-valence excitations within the
\emph{same principal quantum number (n)-shell}, like $4d$-$4f$
transitions in RE, or $5d$-$5f$ transitions in the actinides, pose a
problem because the CVM spread is $\sim$ 20-25~eV, often much larger
than the average core-hole valence-electron attractive potential
($Q$) itself! This places higher lying terms up in the conduction
band, `autoionizing' the extra $f$ electron. This mixing with
continua gives rise to very broad, virtual-bound (V-B)
\emph{Fano-resonances}~\cite{Fano}, that cannot be interpreted in
terms of local models. We note that this is an additional effect to
those involving the decay of the core-hole itself, which have been
elegantly described~\cite{Kotani-JPSJ,No_s-CK_for_5f0} for these
giant dipole resonances (GDR)~\cite{Wendin}.

The strong hybridization of the $5f$ electrons with conduction band
states requires detailed information about interatomic interactions
and band structure effects, ruling out the usefulness of atomistic
models. In this paper, we emphasize that the structure of core-level
excitations within the same $n$-shell, is fundamentally different
from that between different $n$-shells, with the example of the
$5d$$\to$$5f$ \emph{non-resonant inelastic x-ray scattering} (NIXS)
in the actinides. We also show that the high-multipole (HM)
transitions in NIXS to strongly bound CVM states~\cite{Gordon-EPL},
unlike the dipole restricted transitions in XAS and EELS, can still
be treated within local models, but with strongly renormalized
parameters, due to very large \emph{configuration interaction} (CI)
in the final state.

Unlike dipole restricted XAS, NIXS can access transitions involving
high-order multipoles as exemplified by the observation of $d$-$d$
transitions in TM compounds~\cite{Larson,Maurits-PRL}. It also gives
us more information of what the true ground state of the system
actually was, even for dipole-allowed
transitions~\cite{Gordon-EPL,Gordon-MnO}. NIXS uses the first order
scattering, off-resonance, due to the
$(e^{2}/2mc^{2})\vec{A}\cdot\vec{A}$ term in the light-matter
coupling~\cite{Schulke}. The corresponding double differential cross
section is given by~\cite{Schulke} the \emph{Thompson scattering
cross-section}, times the material dependent \emph{dynamical
structure factor} :
\begin{equation}
S(\vec{q},\omega) = \sum_{f}|\langle
f|e^{i\vec{q}\cdot\vec{r}}|i\rangle|^{2}\delta(E_{f}-E_{i}-\hbar\omega)
\end{equation}
where, $\vec{q}$$=$$\vec{k}_{i}$$-$$\vec{k}_{f}$ is the \emph{photon
momentum transfer}, and
$\hbar\omega$$=$$\hbar\omega_{i}$$-$$\hbar\omega_{f}$ is the
\emph{energy loss}. The transition operator can be multipole
expanded~\cite{Maurits-PRL} as : $e^{i\vec{q}\cdot\vec{r}} =
\sum_{l=0}^{\infty}\sum_{m=-l}^{l}i^{l}(2l+1)j_{k}(qr)
C^{(l)*}_{m}(\theta_{\vec{q}},\phi_{\vec{q}})C^{(l)}_{m}(\theta_{\vec{r}},\phi_{\vec{r}})$,
where, $j_{l}(qr)$ are spherical Bessel functions, while
$C^{(l)}_{m}(\theta,\phi)$ are renormalized spherical harmonics,
both of order $l$. Only terms with $|l_{f}-l_{i}|\leq l \leq (l_{f}
+ l_{i})$ and $(l+l_{i}+l_{f})$ \emph{even}, survive in the infinite
sum.

As a relevant illustration we show, in the \emph{inset}-1 to Fig.
1(a), the radial transition probabilities for the various allowed
channels in the $5d$$\to$$5f$ NIXS of the Th$^{4+}$ ($5f^{0}$)
system (relevant for ThO$_{2}$), plotted against $q$, as obtained
using Cowan's atomic Hartree-Fock (H-F) code~\cite{Cowan}. From the
selection rules, this transition has allowed channels for the
\emph{dipole} ($l$$=$$1$), \emph{octupole} ($l$$=$$3$) and the
\emph{triakontadipole} ($l$$=$$5$) sectors. The $q$ and $l$
dependence of the transition probabilities can be understood from
the properties of $j_{l}(qr)$, as described by Haverkort \emph{et
al.}~\cite{Maurits-PRL}.

\begin{figure}
\begin{center}
\includegraphics[angle=0, width=0.45\textwidth]{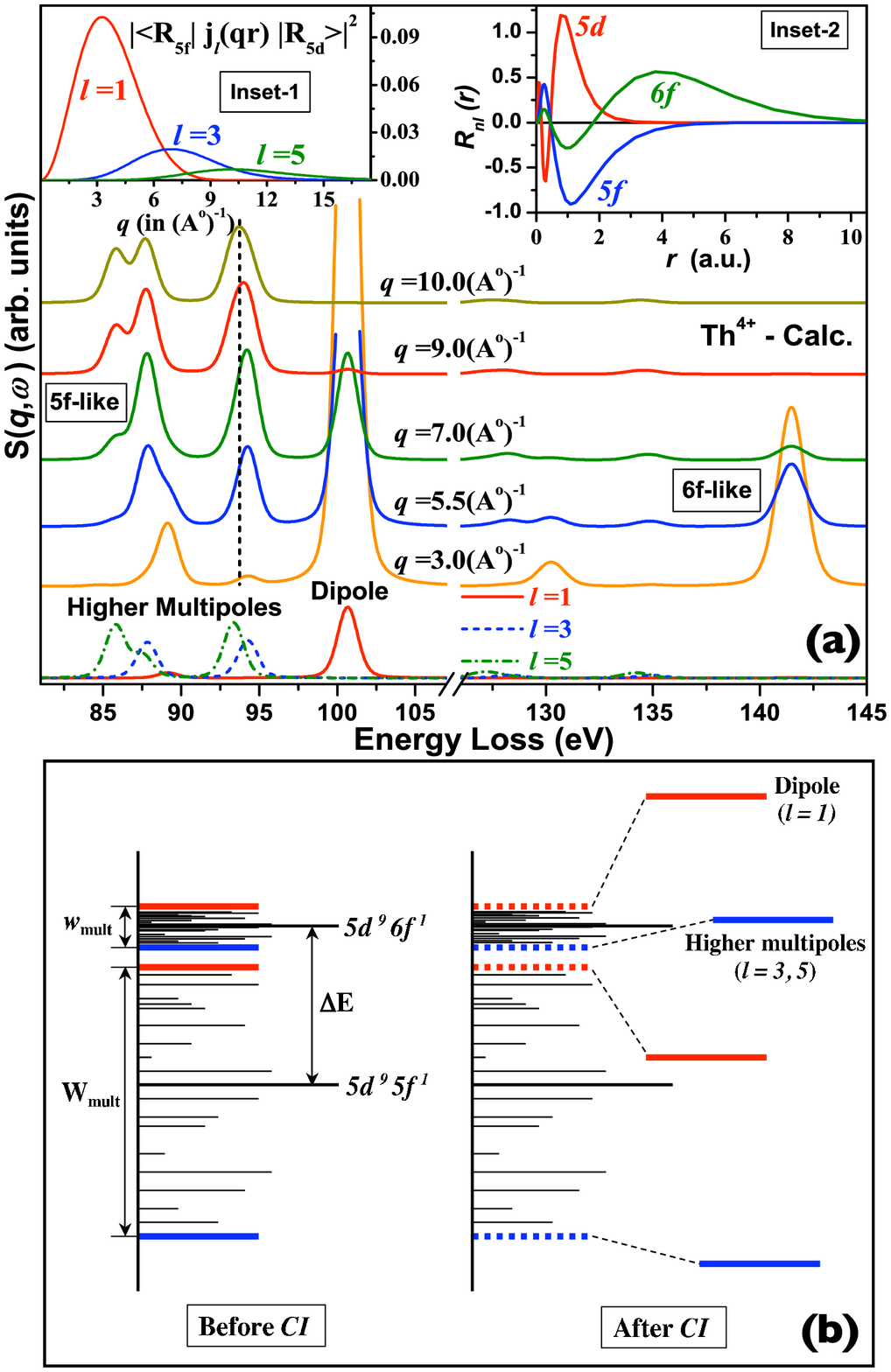}
\caption {(\emph{color online}) (a) The variation of the radial
transition probabilities with $q$, for the three component channels,
$l=1$(\emph{dipole}), $3$(\emph{octupole}) and
$5$(\emph{triakontadipole}) (\emph{inset}-1); plots of the $5d$,
$5f$ and $6f$ atomic H-F radial wavefunctions for Th$^{4+}$
(\emph{inset}-2); and calculated NIXS ($S(q,\omega)$) for Th$^{4+}$
($5f^{0}$) including final-state CI with the $6f$ level, for 60\% of
atomic $CI$($5f$-$6f$) values (\emph{main}). Both the $5f$-like and
$6f$-like regions are shown. The bare component ($l$$=$$1,3,5$)
spectra (\emph{not to scale}) are shown at the bottom. (b) Schematic
illustrating the high sensitivity of the dipole term vis-a-vis the
HM terms, to the strongly term-dependent CI with the $6f$.}
\label{Fig1}
\end{center}
\vspace{-0.75cm}
\end{figure}

While, details of the experimental data are discussed in the
experimental paper~\cite{Bradley-PRL-unpub}, here we briefly remind
the reader of the key results obtained therein, \emph{viz.}~: $(i)$
NIXS data for actinides show the dichotomy that HM features are
sharp and excitonic, while the Fano-like asymmetric~\cite{Fano}
dipole feature is spread out over a very large energy range; $(ii)$
in stark contrast, the calculated low-$q$ dipolar NIXS spectra are
much sharper, intense and similar to the lower energy HM multiplets;
$(iii)$ in order to obtain reasonable agreement with experimental
spectra in terms of the peak positions, we needed to drastically
scale down the $5d$-$5f$ Coulomb and exchange Slater integrals
($F^{k}_{df}$,$G^{k}_{df}$), to 60\% (ThO$_{2}$) or 50\% (UO$_{2}$)
of their atomic H-F values! This is hard to justify for these rather
atomic-like $5f$ wavefunctions~\cite{antonides,ogasawara}. In
passing, we note that the dichotomy, discussed in $(i)$ above, is
also observed in the $4d$$\to$$4f$ NIXS of the RE~\cite{Gordon-EPL}
and the $3p$$\to$$3d$ NIXS of TM compounds~\cite{Gordon-MnO}, and
between the ``doubly forbidden" pre-edge peak and the GDR, in the
O$_{45}$ XAS of actindes~\cite{RMP-Moore}, showing that this is a
generic feature for shallow core-valence transitions within the
\emph{same $n$-shell}.

To understand the origin of the apparent large reductions in the
Slater integrals, we plot in the \emph{inset}-2 of Fig. 1(a), the
$5d$, $5f$ and the $6f$ radial wavefunctions, obtained from an
atomic H-F calculation for ThO$_{2}$ (Th$^{4+}$, ground
configuration $5f^{0}$). The $5d$ and $5f$ orbitals (same $n$-shell)
overlap very strongly, which accounts for the large values of the
atomic ($F^{k}_{df}$,$G^{k}_{df}$) integrals. In contrast, the more
diffuse $6f$ orbital, with an additional radial node, overlaps only
weakly with the $5d$. Hence, a final state (CI) between
$5d^{9}5f^{1}$ and $5d^{9}6f^{1}$, \emph{via} the $CI$($5f$-$6f$)
matrix elements~\cite{Rkdefn}, would effectively expand the radial
part of the $5f$ wavefunction, reducing the $5d$-$5f$ Slater
integrals. Now, the mixing depends also on the energy separation,
$\Delta E$, between the center-of-gravities (CG) of configurations
involved, which is smaller in the actinides, than between
$4d^{9}4f^{1}$ and $4d^{9}5f^{1}$ in the RE. This results in a
strong multiplet-dependent CI, because the multiplet splitting in
$5d^{9}5f^{1}$ is much larger than that in $5d^{9}6f^{1}$, as shown
schematically in Fig. 1(b).
To illustrate this point, the NIXS spectra for ThO$_{2}$ (Th$^{4+}$)
are calculated (using the XTLS8.3 code~\cite{Tanaka-XTLS}) on the
basis of the above model, as a coherent combination of the
transitions $5f^{0}$$\to$$5d^{9}5f^{1}$ and
$5f^{0}$$\to$$5d^{9}6f^{1}$~\cite{Longer-paper}. Here we leave the
($F^{k}_{df}$,$G^{k}_{df}$) integrals unaltered at their atomic H-F
values, while the scaling of $CI$($5f$-$6f$) is varied to obtain
agreement with experimental peak positions. This presents a more
natural and physical mechanism for understanding the reduced
multiplet spread observed experimentally. The fact that we need to
scale the CI integrals merely indicates that the band-like $6f$
state is poorly approximated by the atomic H-F calculations, and
that CI with numerous other states is neglected here. The full
$q$-dependent spectra, at the optimized 60\% scaling of
$CI$($5f$-$6f$)(Fig. 1(a)), is spread over a wide energy range and
consists of \emph{$5f$-like} and\emph{ $6f$-like} regions (marked in
the figure), although most of the spectral weight lies in the
$5f$-like region due to dominance of radial matrix elements. The CI
serves to reduce the $5d$-$5f$ multiplet spread considerably (while
enhancing the $5d$-$6f$ spread), especially pushing the dipole peak
close to the HM peaks, in good agreement with
experiments~\cite{Bradley-PRL-unpub}. Also the component spectra for
the $l$$=$$1,3,5$ channels plotted at the base of Fig. 1(a)
(vertical scale not to be compared with the actual spectra above)
clearly show that it also reproduces the mild shift of the HM peak
at $\sim$ 93 eV, due to a $q$-dependent weight transfer between the
$l$$=$$3$ and $l$$=$$5$ channels, as seen in
experiments~\cite{Bradley-PRL-unpub}. The dipole states, in general,
mix much more and respond much more sensitively to the CI than the
HM states. This also means that there is a substantial amount of
interference between the $5f$-like and $6f$-like dipole transitions.

\begin{figure}
\begin{center}
\includegraphics[width=0.40\textwidth]{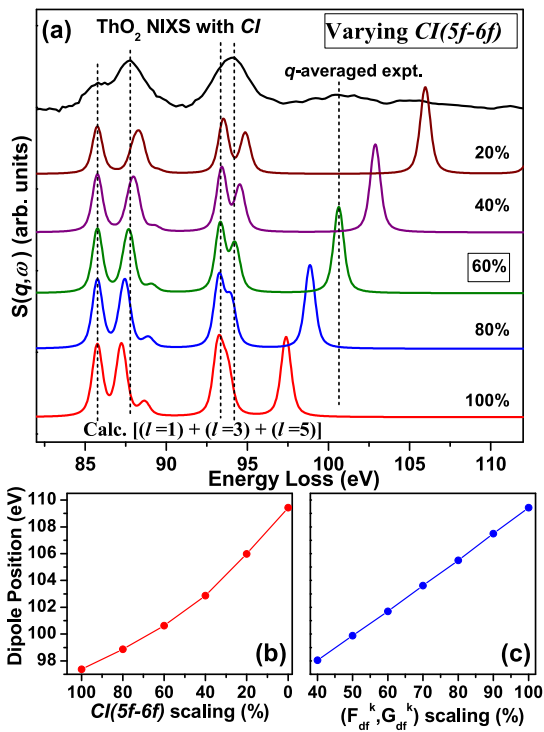}
\vspace{-0.2cm}
\caption {(\emph{color online}) (a) Experimental
$q$-averaged NIXS for ThO$_{2}$, compared with the component
($l$$=$$1,3,5$) summed calculated spectra for atomic values of
($F^{k}_{df}$,$G^{k}_{df}$), and the $CI$($5f$-$6f$) varied from a
scaling of 20\% to their full atomic values. \emph{Only
peak-positions are relevant.} The best agreement is obtained at 60\%
scaling of $CI$($5f$-$6f$). Variation of the high-energy dipole peak
position with varying scaling of : (b) $CI$($5f$-$6f$), and (c)
($F^{k}_{df}$,$G^{k}_{df}$). While the latter shows a simple linear
trend, the former shows a parabolic behavior with signs of
saturation.} \label{Fig2}
\end{center}
\vspace{-0.75cm}
\end{figure}

The origin of the higher sensitivity of the dipole compared to the
HM is explained schematically in Fig. 1(b). Before CI (\emph{left}),
the difference ($\Delta E$) in the CG energies for the two
multiplets, $5d^{9}5f^{1}$ (width $W_{mult}$) and $5d^{9}6f^{1}$
(width $w_{mult}$), is $\sim$ 23-24 eV, from H-F calculations.
Although both the multiplets involve exactly the same terms,
$W_{mult}$ ($\sim$ 25 eV) is much larger than  $w_{mult}$($\sim$ 7
eV), demonstrating the difference between states involving the
\emph{same} versus \emph{different principle quantum numbers}. Also,
for a less-than-half-filled system with a low $J$ ground state
($J$$=$$0$ for ThO$_{2}$) , the highest multiplets (\emph{red}) are
generally \emph{dipole-allowed}, while the lowest ones (\emph{blue})
are the \emph{HM-allowed} terms. Now $CI$($5f$-$6f$) only mixes
terms of the same symmetry~\cite{Cowan}, \emph{e.g.}, the \emph{red}
(\emph{blue}) states at the \emph{top} (\emph{bottom}) of
$5d^{9}5f^{1}$, mix only with the \emph{red} (\emph{blue}) states at
the \emph{top} (\emph{bottom}) of $5d^{9}6f^{1}$. The result after
mixing is shown in the \emph{right panel}. Since $w_{mult} <<
W_{mult}$, the effective energy denominator for mixing of the
\emph{red (dipole)} terms is much smaller than that for the
\emph{blue (HM)} terms, causing the observed differences in the
shifts. Due to these strong correlation effects, the effective
``screening" becomes highly term dependent, explaining the strong
asymmetry in the behavior of the dipole and the HM terms, which is
not captured by a uniform reduction of the Slater
integrals~\cite{Bradley-PRL-unpub}.

To show that the two approaches are qualitatively different, we
compare in Fig. 2(a) the experimental $q$-averaged NIXS
(\emph{topmost}), with the calculated sum of the three component
spectra ($l$$=$$1,3$ and $5$), keeping $(F^{k}_{df},G^{k}_{df})$
fixed at their atomic values, while the $CI$($5f$-$6f$) are switched
on and gradually increased to their atomic value (100\%) (\emph{top
to bottom}). For this purpose only the peak positions are relevant.
As already noted, a good agreement with HM peak positions is
obtained for 60\% reduction of $CI$($5f$-$6f$). But more
importantly, with gradual uniform reduction in
($F^{k}_{df}$,$G^{k}_{df}$) we would expect a linear movement of the
dipole towards the HM features. On the other hand with changing
degree of CI, the dipole moves towards the HM peaks in a nonlinear
manner, showing signs of saturation, as governed by
\emph{level-repulsion physics}. This contrasting behavior is shown
in Figs. 2(b) and 2(c), where we have plotted the dipole peak
position in the two cases, as a function of the scaling of
($F^{k}_{df}$,$G^{k}_{df}$) and of $CI$($5f$-$6f$), respectively.

\begin{figure}
\begin{center}
\includegraphics[angle=0,width=0.45\textwidth]{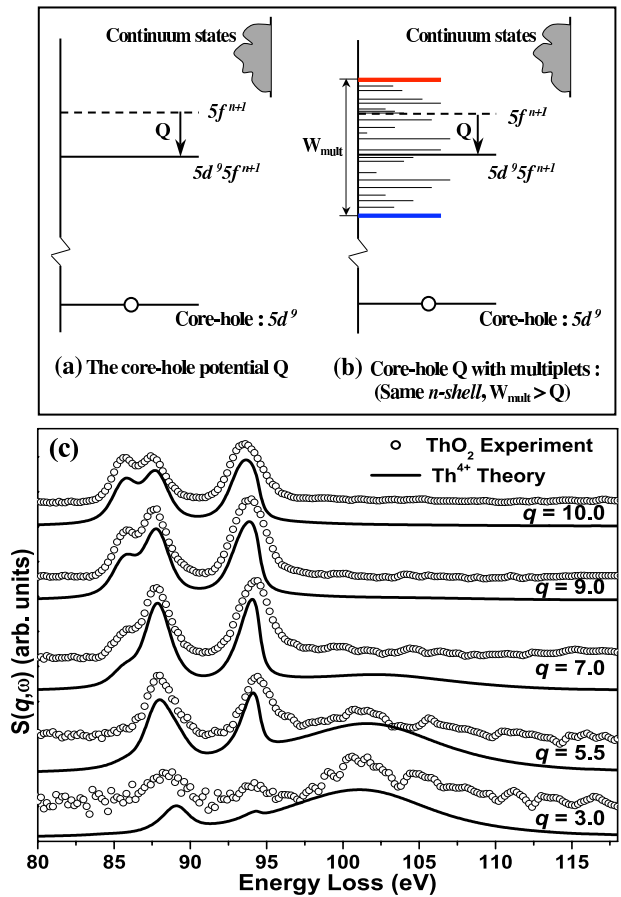}
\caption {(\emph{color online}) Schematics showing (a) the effect of
a scaler core-hole potential ($Q$) on the NIXS final state; and
(b)its fanning out due to strong core-valence multiplet effects
leading to a gradual change from bound-states (lower terms) to V-B
resonances (higher terms) as a function of energy; (c) the
$S(q,\omega)$ from a model calculation for Th$^{4+}$(\emph{lines})
including a Fano-effect of the mixing of the atomic $5d$-$5f$
transition with that to a fictitious broad band, causing a broad,
asymmetric lineshape of the high energy dipole term, that agrees
well with experimental $q$-dependent NIXS for
ThO$_{2}$~\cite{Bradley-PRL-unpub} (symbols).}\label{Fig3}
\end{center}
\vspace{-0.75cm}
\end{figure}

We now turn to the problem of the experimentally observed low
relative amplitude, large width and non-lorentzian line shape of the
GDR. This can be understood on the basis of Figs. 3(a)-(b).  Fig.
3(a) shows the NIXS final state, $5d^{9}5f^{n+1}$, of a $5f^{n}$
actinide system. In the absence of the $5d$ core-hole, this would be
identical to the inverse photoemission final state $5f^{n+1}$
(dashed line), and lies just below or within a continuum (shaded).
Primarily, the core-hole provides an attractive scalar potential,
$Q$, which is often large enough to pull down the $5d^{9}5f^{n+1}$
state, out of the continuum, forming a bound \emph{core-hole
exciton}. However, in a more complete picture (Fig. 3(b)), the
core-hole $f$-electron multipole interaction, also yields a very
broad multiplet structure (width $W_{mult}$) about the CG of
$5d^{9}5f^{n+1}$. The Slater integrals are very large for these
transitions occurring within the same $n$-shell, implying
$W_{mult}$$>>$$Q$. Thus while the HM states towards the bottom of
the multiplet, still form bound states, the high-lying dipolar terms
are pushed up into the continuum, offsetting the effect of $Q$, and
their mixing gives rise to the GDR with characteristic Fano
lineshapes, as seen experimentally. This physical picture clearly
demonstrates why the dipole and the HM states show a crossover from
V-B to bound character, in the $5d$$\to$$5f$
(actinides)~\cite{Bradley-PRL-unpub}, the $4d$$\to$$4f$  (RE
compounds)~\cite{Gordon-EPL}, or the $3p$$\to$$3d$ (TM compounds)
edges~\cite{Gordon-MnO}.

In Fig. 3(c) we show $S(q,\omega)$ from a model calculation for
Th$^{4+}$ (lines) that includes transitions from the $5d$ core-level
to both the $(5f,6f)$ levels as before (\emph{allowed}
$l$$=$$1,3,5$), and to a fictitious $7p$-like discretized band
(\emph{allowed} $l$$=$$1,3$). The band is so positioned that it
starts below the high energy dipole state, but above the HM states.
The $l$$=$$1,3$ channels can interfere \emph{via} the
$CI$($5f$-$7p$) matrix elements~\cite{Rkdefn, Longer-paper}. We find
that while the lower lying $l$$=3,5$ peaks remain sharp and
excitonic, the high-energy dipole feature forms a GDR, just as
discussed above. Interestingly, a dipole-allowed peak present at
lower energy ($\sim$ 88 eV) is not broadened by this mechanism,
implying that \emph{the position within the multiplet, rather than
the symmetry of the state, decides its fate}. A fairly good
comparison with the experimental $q$-dependent NIXS for
ThO$_{2}$~\cite{Bradley-PRL-unpub} (symbols) is obtained if we use a
somewhat larger Lorentzian width for the GDR than the HM states, in
order to simulate the multiplet dependent core-hole decay
probabilities, not included in the present calculation. It is
important to note that this dichotomy, between the dipole and the
HM, could be reversed in cases where the dipole allowed states are
lower in energy than the HM, like for more-than-half-filled systems
which have large ground state $J$ values~\cite{Longer-paper}.

In conclusion, the modeling of \emph{same $n$-shell} NIXS is
complicated by the simultaneous presence of V-B and bound states
within the same final-state multiplet. The complex V-B resonances,
involving non-local effects, provide insight into the hybridization
of the locally excited core-electron with continua, and about
core-hole decay processes. The dipole-forbidden bound states (not
prominent in XAS) are modeled using a local, atomic CI approach,
that provides direct ground state information. It also explains the
apparent strong reduction of the atomic Slater integrals, which
effectively become term dependent. The HM transitions and especially
their angular dependence in single crystal
studies~\cite{Larson,Maurits-PRL} are also expected to provide
detailed information on the importance of ``orbital-ordering" in the
ground state and changes at phase transitions in ``hidden order"
materials like URu$_{2}$Si$_{2}$~\cite{URu2Si2}. The V-B resonances
are modeled with an additional term, mixing the $5f$ states with a
conduction band continuum, resulting in their broad and Fano-like
line shapes. More realistic approaches to the latter would include
an energy-dependent hybridization with a realistic density of
states, and explicit core-hole decay processes. This is a topic of
future investigations.

SSG and GAS acknowledge funding from the Canadian agencies NSERC,
CFI and CIFAR. GTS and JAB acknowledge support from University of
Washington, and the U.S. Department of Energy.


\begin{thebibliography}{99}
\bibitem{No_s-CK_for_5f0} It is important to note that such a
mechanism, relying on the coupling of the $5d$ photoexcited state
to the $5f$ PES continuum \emph{via} super-Coster-Kronig channels,
is not operative for a $5f^{0}$ system (ThO$_{2})$~\cite{Sonntag-1},
as discussed here.
\bibitem{Rkdefn} In our case, the $CI$($\alpha$-$\beta$) matrix elements refer to the Slater Coulomb and exchange integrals $R^{k}(5d,\alpha,5d,\beta)$.
\end{thebibliography}
\end{document}